\documentclass{pinchcr}   

\usepackage{graphicx}
\graphicspath{{./FIGFILES/}}

\usepackage{latexsym}
\usepackage{amsmath}
\usepackage{amssymb}

\title{Schr\"odinger Cat States in Circuit QED}

\subtitle{Lectures presented at the Les Houches Summer School, Session CVII--Current Trends in Atomic Physics, July 2016.}

\author{S. M. Girvin}
\affiliation{Yale Quantum Institute\\PO Box 208334\\New Haven, CT 06520-8263 USA}
\authors{1}

\begin{document}

\maketitle


\preface
The last 15 years have seen spectacular experimental progress in our ability to create, control and measure the quantum states of superconducting `artificial atoms' (qubits) and microwave photons stored in resonators.   In addition to being a novel testbed for studying strong-coupling quantum electrodynamics in a radically new regime, `circuit QED,' defines a fundamental architecture for the creation of a quantum computer based on integrated circuits with semiconductors replaced by superconductors.  The artificial atoms are based on the Josephson tunnel junction and their relatively large size ($\sim$mm) means that the couple extremely strongly to individual microwave photons.  This strong coupling yields very powerful state-manipulation and measurement capabilities, including the ability to create extremely large ($>100$ photon) `cat' states and easily measure novel quantities such as the photon number parity.   These new capabilities have enabled a highly successful scheme for quantum error correction based on encoding quantum information in Schr\"odinger cat states of photons.

\acknowledgements

The ideas described here represent the collaborative efforts of many students, postdocs and faculty colleagues who have been members of the Yale quantum information team over the past 15 years.   The author is especially grateful for the opportunities he has had to collaborate with long-time friends and colleagues Michel Devoret, Leonid Glazman, Liang Jiang and Rob Schoelkopf as well as frequent visitor, Mazyar Mirrahimi.  Most of the ideas presented in these notes originated primarily with them and not the author.  This work was supported by the National Science Foundation through grant DMR-1609326, by the Army Research Office and the Laboratory of Physical Sciences through grant ARO W911NF1410011, and by the Yale Center for Research Computing and the Yale Quantum Institute.

\tableofcontents

\maintext



\chapter{Schr\"odinger Cat States in Circuit QED}

\section{Introduction to Circuit QED}

Circuit quantum electrodynamics (`circuit QED') \shortcite{BlaisCQEDtheory2004,WallraffCQED2004,Blais2007,WiringUpQuantumSystems,LesHouchesQuantumMachines,%
Devoret-Schoelkopf-Science-Review-2013,NiggBlackBox,MultiportBBQ_Divencenzo,Vool_Devoret_QCircuit_Review} describes the quantum mechanics and quantum field theory of superconducting electrical circuits operating in the microwave regime near absolute zero temperature.  It is the analog of cavity QED in quantum optics with the role of the atoms being played by superconducting qubits.  A detailed pedagogical introduction to the subject with many references is available in the author's lecture notes from the 2011 Les Houches School on Quantum Machines \shortcite{LesHouchesQuantumMachines}.  The present notes will therefore provide only a brief introductory review of the subject and will focus primarily on novel quantum states that can be produced using the strong coupling between the artificial atom and one or more cavities.

It is a basic fact of QED that each normal mode of the electromagnetic field is an independent harmonic oscillator with the quanta of each oscillator being photons in the corresponding mode.   For an optical or microwave cavity, such modes are standing waves trapped inside the cavity.  Cavities typically contain many-modes, but for simplicity we will focus here on the case of a single mode, coupled to an artificial atom approximated as having only two-levels (and hence described as a pseudo spin-1/2).  This leads to the Jaynes-Cummings model \shortcite{BlaisCQEDtheory2004,LesHouchesQuantumMachines}
\begin{equation}
H=\tilde\omega_\mathrm{c} a^\dagger a + \frac{\tilde\omega_\mathrm{q}}{2} \sigma^z + g [a\sigma^+ + a^\dagger \sigma^-]
\end{equation}
describing the exchange of energy between the cavity and the atom via photon absorption and emission within the RWA (rotating wave approximation).  Here $\tilde\omega_\mathrm{c}$ is the (bare) cavity frequency, $\tilde\omega_\mathrm{q}$ is the (bare) qubit transition frequency and $g$ is the vacuum Rabi coupling which, because of the large qubit size and small cavity volume can be enormously large ($\sim 150$MHz which is several percent of the $\sim5$GHz qubit frequency and several orders of magnitude larger than the cavity and qubit decay rates).

\begin{figure}[htb]        
\begin{center}              
 \includegraphics[width=4.0in]{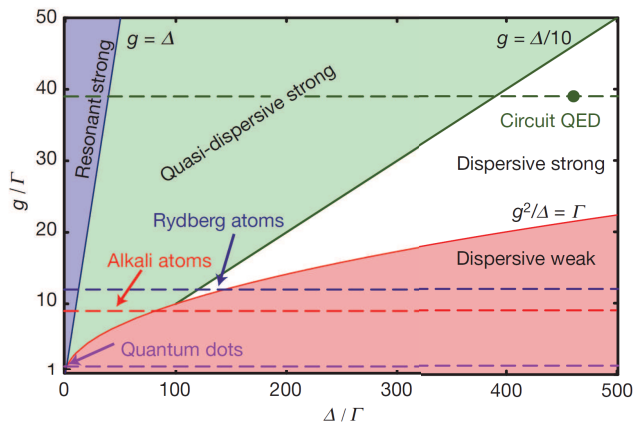}
\end{center}
\caption{A phase diagram for cavity QED. The parameter space is described by the atom-photon coupling strength, $g$, and the
detuning $\Delta$ between the atom and cavity frequencies, normalized to the rates of decay represented by
$\Gamma=\max\left[\gamma,\kappa,1/T\right]$. Here $\gamma$ and $\kappa$ are the qubit and cavity decay rates respectively, and $T$ is the transit time for atoms passing through the cavity. Different cavity QED systems, including Rydberg atoms, alkali atoms, quantum dots, and
circuit QED, are represented by dashed horizontal lines.  Since the time this graph was first constructed $\Gamma$ has decreased dramatically putting circuit QED systems much deeper into the strong dispersive regime. Adapted from (Schuster et al., 2007).
}
\label{fig:StrongDispersiveSchuster2007}    
\end{figure}

A `phase diagram' \shortcite{Schuster2007} for the Jaynes-Cummings model for different realizations of cavity and circuit QED  is illustrated in Fig.~(\ref{fig:StrongDispersiveSchuster2007}). The vertical axis represents the strength $g$ of the atom-photon coupling in the cavity and the horizontal axis represents the detuning $\Delta = \tilde\omega_\mathrm{q}-\tilde\omega_\mathrm{c}$ between the atomic transition and the cavity frequency.  Both are expressed in units of $\Gamma=\max\left[\gamma,\kappa,1/T\right]$, where $\gamma$ and $\kappa$ are the qubit and cavity decay rates respectively, and for the case of (`real') atoms, $T$ is the transit time for the atoms passing through the cavity\footnote{Another benefit of circuit QED is that our artificial atoms are `glued down' and stay in the cavity indefinitely.}   In the
region $g\geq\Delta$ the qubit and cavity are close to resonant and first-order degenerate perturbation theory applies. The lowest two excited eigenstates are the upper and lower `polaritons' which are coherent superpositions of atom+photon excitations \shortcite{BlaisCQEDtheory2004,LesHouchesQuantumMachines}.  They are essentially the bonding and anti-bonding combinations of `atom $\pm$ photon.' An excited state atom introduced into the cavity will `Rabi flop' coherently between being an atomic excitation and a photon at the `vacuum Rabi rate' $2g$ (which is the energy splitting between the upper and lower polariton). Within this degenerate region there is a `strong-coupling' regime $g\gg\Gamma$, in which the cavity and qubit undergo multiple vacuum Rabi oscillations prior to decay.

We will focus here on the case where the detuning $\Delta$ between the qubit and the cavity is large ($\Delta\gg g$).  Because of the mismatch in frequencies, the qubit can only virtually exchange energy with the cavity and the vacuum Rabi coupling can be treated in second-order perturbation theory.  Applying a unitary transformation which eliminates the first-order effects of $g$ and keeping only terms up to second-order yields \shortcite{BlaisCQEDtheory2004,LesHouchesQuantumMachines}
\begin{equation}
H=\omega_\mathrm{c} a^\dagger a + \omega_\mathrm{q}|e\rangle\langle e| - \chi a^\dagger a |e\rangle\langle e|,
\end{equation}
where we have shifted the overall energy by an irrelevant constant, $\omega_\mathrm{c}$ and $\omega_\mathrm{q}$ are renormalized cavity and qubit frequencies and $|e\rangle\langle e|$ is the projector onto the excited state of the qubit.  The quantity $\chi\approx \frac{g^2}{\Delta}\sim 2\pi*(2-10\mathrm{MHz})$ is the effective second-order coupling known as the dispersive shift.\footnote{The simple expression given here for the dispersive shift is not accurate for the so-called transmon qubit because it is a weakly anharmonic oscillator and not necessarily well-approximated as a two-level system.  For a quantitatively more accurate approach see \shortcite{NiggBlackBox,LesHouchesQuantumMachines}.}  In the dispersive regime, the qubit acts like a dielectric whose dielectric constant depends on the state of the qubit.  This causes the cavity frequency to depend on the state of the qubit.

\begin{figure}[hbt]        
\begin{center}              
 \includegraphics[width=4.0in]{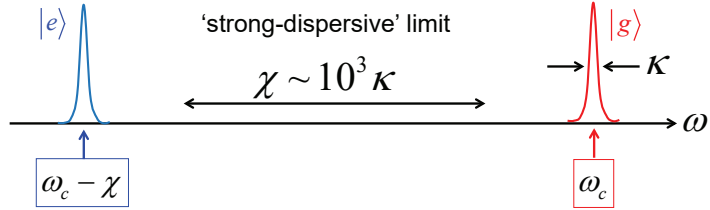}
\end{center}
\caption{Illustration of cavity response (susceptibility) depending on the state of the qubit.  In the qubit ground state, the cavity resonance frequency is $\omega_\mathrm{c}$.  When the qubit is in the excited state, the cavity frequency shifts by $-\chi$, which can be thousands of times larger than the cavity linewidth $\kappa$.  This is the `strong-dispersive' regime. When the dispersive shift is less than a linewidth of the cavity (or the qubit) we are in the `weak-dispersive' regime.
}
\label{fig:StrongDispersiveCavityShift}    
\end{figure}

As illustrated in Fig.~\ref{fig:StrongDispersiveCavityShift}, when the qubit is in the excited state, the cavity frequency shifts by $-\chi$.  As we will soon discuss, this dispersive shift has a number of useful applications, but the simplest is that it can be used to read out the state of the qubit \shortcite{BlaisCQEDtheory2004}.  We can find the qubit state by measuring the cavity frequency by simply reflecting microwaves from it and measuring the resulting phase shift of the signal.  Because the cavity frequency is very different from the qubit frequency, the photons in the readout do not excite or deexcite the qubit and the operation is non-destructive. Even though $\chi$ is a second-order effect in $g$ it still can be several thousand times larger than the respective cavity and qubit decay rates, $\kappa$ and $\gamma$.  This is the so-called `strong-dispersive' regime \shortcite{Schuster2007,LesHouchesQuantumMachines} of the phase diagram.  The ability to easily enter this regime gives circuit QED great advantages over ordinary quantum optics and will prove highly advantageous for quantum state manipulation, control and error correction.

\begin{figure}[hbt]        
\begin{center}              
 \includegraphics[width=3.0in]{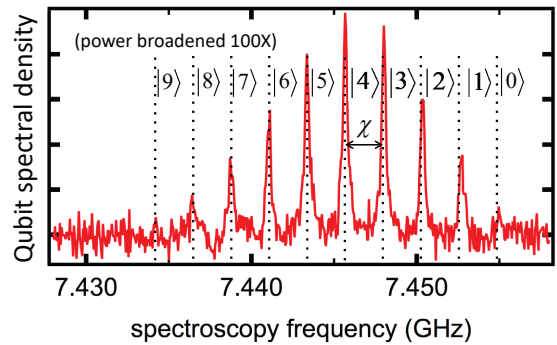}
\end{center}
\caption{Quantum jump spectroscopy of a transmon qubit dispersively coupled to a cavity illustrating the quantized light shift of the qubit transition frequency depending on how many photons are in the cavity.  The weakly damped cavity is driven to produce a coherent state whose time-averaged photon number distribution is Poisson.  A weak spectroscopy tone is used to excite the qubit at different frequencies.   The spectral density of the qubit is determined by the probability that the spectroscopy tone excites the qubit.  In this case a relatively strong spectroscopy tone was used, resulting in a power-broadening of the spectral line by approximately 100x.  The spectral peaks are actually about $10^3$ times narrower than their separation.  The symbol $|n\rangle$ denotes the number of photons in the cavity associated with each spectral peak of the qubit.  The quantization of the light shift clearly shows that microwaves are particles!  Courtesy R.~Schoelkopf group.
}
\label{fig:QuantizedLightShift}    
\end{figure}

The very same dispersive coupling term can also be viewed as producing a quantized `light shift' of the qubit transition frequency by an amount $-\chi$ for each photon that is added to the cavity.  Fig.~\ref{fig:QuantizedLightShift} shows quantum jump spectroscopy data illustrating this quantized light shift.
In ordinary spectroscopy of a medium (a gas, say) one measures the spectrum by the absorption of the spectroscopic light at different frequencies.  Here we instead use quantum jump spectroscopy in which we use the dispersive readout of the state of the qubit to determine when the qubit jumps to the excited state as a result of excitation by the spectroscopy tone.

The dispersive term commutes with both cavity photon number and qubit excitation number.  It is thus `doubly QND,' and can be utilized to make quantum non-demolition measurements of the cavity (using the qubit) and the qubit (using the cavity).  In the `strong-dispersive' region $\chi\gg\Gamma$, quantum non-demolition measurements of photon number are possible.  We will see later that in this regime it is even possible to make a QND measurement of the photon number parity without learning the value of the photon number!

In the `weak-dispersive regime' ($\chi\ll\Gamma$) of the phase diagram, the light shift per photon is too small to dispersively resolve individual photons, but a QND measurement of the qubit can still be
realized by using many photons. The qubit-dependent cavity frequency shift is less than the linewidth of the cavity but with enough photons over a long enough period, this small frequency shift can be detected (using the small phase shift of the reflected signal, assuming that the qubit lifetime is longer than the required mesaurement time) \shortcite{Clerk2008}.

\section{Measurement of Photon Number Parity}
In the strong-dispersive regime, if we measure the qubit transition frequency the state of the cavity collapses to a Fock state with definite photon number determined by the measured light-shift of the qubit transition frequency. Unlike a photomultiplier, this measurement is not only photon-number resolving, it is QND.  The photon is not absorbed and we can repeat the measurement many times to overcome any imperfections in the measurement (improving the quantum efficiency and lowering the dark count).  This novel feature has the potential to dramatically accelerate searches for axion dark matter particles which convert into microwave photons in the presence of a strong magnetic field \shortcite{AxionSearchQTech}.

Another extremely powerful feature of the strong-dispersive regime is that it gives us the ability to measure the parity of the photon number without learning the photon number itself. Why is this interesting?  It turns out that in quantum systems, what we do \emph{not} measure is just as important as what we \emph{do} measure!  One way to measure the parity would be to measure the eigenvalue of the photon number operator $\hat n$.  If the result is $m$, then we assign parity $\Pi=(-1)^m$.  This process works, but it does a great deal of `damage' to the state by collapsing it to the Fock state $|m\rangle$. This particular measurement has strong `back action.'  Let us formalize this by considering an arbitrary cavity state $|\psi\rangle$.  The state of the system conditioned on the measurement result
\begin{equation}
|\psi\rangle_m =  \frac{|m\rangle\langle m|\psi\rangle}{[\langle\psi|m\rangle\langle m|\psi\rangle]^{\frac{1}{2}}}=|m\rangle
\end{equation}
is completely independent of the starting state.  (Only the probability of obtaining the measurement outcome $m$ depends on $|\psi\rangle$.)

If we somehow had a way to directly measure the eigenvalue of the photon number parity operator
\begin{equation}
\hat \Pi= e^{i\pi a^\dagger a}=(-1)^{\hat n},
\label{eq:parity_op}
\end{equation}
the back action would be much weaker.  This is because there are only two measurement outcomes and the projection onto the even or odd subspaces only cuts out half of the overall Hilbert space, rather than all but one Fock state.\footnote{More formally, the projector onto Fock state $|m\rangle$ has unit trace, $\mathrm{Tr}\, |m\rangle\langle m|=1$, while the parity-subspace projectors defined in eqns~\ref{eq:Piplus}-\ref{eq:Piminus} have infinite trace, $\mathrm{Tr}\,\hat\Pi_\pm=\mathrm{Tr}\,\hat I=\infty$.  Once could (in principle) map the entire Hilbert space into the even or the odd subspaces! }   Conditioned on the parity measurement being $\pm 1$ we obtain
\begin{equation}
|\psi_\pm\rangle =\frac{\hat \Pi_\pm|\psi\rangle}{[\langle\psi|\hat \Pi_\pm|\psi\rangle]^\frac{1}{2}},
\end{equation}
where
\begin{eqnarray}
\hat \Pi_+&=& \sum_{j\in \mathrm{even}}|j\rangle\langle j|,\label{eq:Piplus}\\
\hat \Pi_-&=& \sum_{j\in \mathrm{odd}}|j\rangle\langle j|.\label{eq:Piminus}
\end{eqnarray}
Because we don't learn the value of the photon number, only its parity, the back action is minimized to that associated with the information we wanted to learn.  In the first method we learned too much information (namely the value of the photon number not just its parity).  The importance of what you do not measure will come to the fore when we discuss how Schr\"odinger cat states can be used for quantum error correction.

The parity operator in eqn~\ref{eq:parity_op} is a non-trivial function of the number operator.  At first sight it appears to be quite difficult to measure.  It turns out however to be very straightforward to create a situation in which we entangle the state of the qubit with the parity of the cavity.  Measuring the qubit then tells us the parity and nothing more.  To see how this works, let us move to the interaction picture (i.e.\ go to an appropriate rotating frame) where the time evolution is governed solely by the dispersive coupling. Moving to a rotating frame via the time-dependent unitary
\begin{equation}
U(t)=e^{iH_0t},
\end{equation}
where
\begin{equation}
H_0=\omega_\mathrm{c} a^\dagger a + \omega_\mathrm{q}|e\rangle\langle e|,
\end{equation}
the new Hamiltonian in the rotating frame becomes
\begin{equation}
V = UHU^\dagger - Ui\frac{d}{dt}U^\dagger=-\chi a^\dagger a |e\rangle\langle e|.
\end{equation}

Now consider time evolution under this Hamiltonian for a period $t=\pi/\chi$.  The unitary evolution operator is
\begin{equation}
U_\pi=e^{+i\pi a^\dagger a |e\rangle\langle e|}.
\end{equation}
Using the fact that $|e\rangle\langle e|$ is a projector, we can reformulate this in terms of the photon number parity operator as
\begin{equation}
U_\pi=|g\rangle\langle g| + \hat \Pi |e\rangle\langle e|=\hat \Pi_+ \hat I + \hat \Pi_- \sigma^z,
\end{equation}
where $|g\rangle\langle g|$ is the projector onto the qubit ground state.  Now sandwich this with Hadamard operators.  These operators interchange the roles of the $x$ and $z$ components of the spin: $H|g\rangle=|+x\rangle$ and $H|e\rangle=|-x\rangle$).  This yields
\begin{equation}
HU_\pi H= \hat \Pi_+ \hat I + \hat \Pi_- \sigma^x.
\label{eq:paritymesprotocol}
\end{equation}
Thus if the photon number is even, nothing happens to the qubit, whereas if the photon number is odd the qubit is flipped.  Hence we have successfully entangled the parity with the qubit state.  Starting with the qubit in the ground state, applying $HU_\pi H$ and then measuring the state of the qubit tells us the parity but not the photon number.

Essentially what we have done is the following.  We use the Hadamard gate to put the spin in the $+x$ direction on the Bloch sphere.  We then allow the qubit to precess around the $z$ axis at a rate that depends on the quantized light shift.  If the parity is even, the precession is through an even integer multiple of $\pi$ radians and returns the qubit to the starting point (and erasing the information on the value of the even integer).  If the parity is odd, the qubit precesses through an odd integer multiple of $\pi$ radians and ends up pointing in the $-x$ direction on the Bloch sphere.  The second Hadamard gate converts $\sigma^x$ to $\sigma^z$ which we then measure to determine the parity of the cavity photon number.

Provided that we are in the strong-dispersive limit (more precisely, provided that $\chi\gg \bar n \kappa$ where $\bar n=\langle\psi|\hat n |\psi\rangle$) then there is very little chance that an error will result from the cavity losing a photon during the course of the parity measurement.  One could also imagine other errors which would make the measurement non-QND.  The manipulations of the qubit could for example accidentally add a photon to the cavity.  This is unlikely however because the qubit and cavity are strongly detuned from each other.  Ref.~\shortcite{Nissim16} were able to make parity measurements of a high-Q storage cavity that were 99.8\% QND and hence could be repeated hundreds of times without extraneous damage to the state.  (The fidelity with which the parity could be determined in a single measurement was 98.5\% due to uncertainties in the readout, but these uncertainties were not associated with any adverse back action on the cavity.)

\section{Application of Parity Measurements to State Tomography}

Quantum computation relies on the ability to create and control complex quantum states.  The dimension of the Hilbert space grows exponentially with the number of qubits and the task of verifying the accuracy of a particular state (state tomography) or verifying a particular transformation of that state (process tomography) becomes exponentially difficult. To fully determine the state, one has to be able to measure not just the states of individual qubits, but also measure non-local multi-qubit correlators of arbitrary weight $\langle \sigma_i^z\sigma_j^z\sigma_k^x\sigma_l^y....\sigma_n^z\rangle$, a task which is quite difficult

  For bosonic states of cavities, the quantum jump spectroscopy illustrated in Fig.~\ref{fig:QuantizedLightShift} gives an excellent way to measure the photon number distribution in the state of the cavity.  This however is far short of the information required to fully specify the quantum state.  We need not just the probabilities of different photon numbers but the probability \emph{amplitudes}.  In general, the state need not be pure (and the processes we are studying need not be unitary) and so we need to be able to measure the full density matrix which has both diagonal and off-diagonal elements.

It turns out that the ability to make high-fidelity measurements of the photon number parity gives an extremely simple and powerful way to measure the Wigner function.  As shown in App.~\ref{app:Wigner}, the Wigner function provides precisely the same information stored in the density matrix but displays it in a very convenient and intuitive format.  The Wigner function $W(X,P)$ is a quasi-probablilty distribution in phase space and can be obtained by the following simple and direct recipe \cite{Davidovich1}:  (1) displace the oscillator in phase space so that the point $(X,P)$ moves to the origin; (2)  then measure the value of the photon number parity.  By repeating many times, one obtains the expectation value of this `displaced parity'
\begin{equation}
W(X,P)=\frac{1}{\pi\hbar}\mathrm{Tr}\,\left\{{\mathcal D}(-X,-P)\rho {\mathcal D}^\dagger(-X,-P) \hat \Pi\right\},
\label{eq:dispPa}
\end{equation}
where $\mathcal D$ is the displacement operator.  As shown in App.~\ref{app:Wigner}, this is precisely the desired Wigner function which fully characterizes the (possibly mixed) state of the system.  Related methods in which one measures not the parity but the full photon number distribution of the displaced state (using quantum jump spectroscopy) can in principle yield even more robust results in the presence of measurement noise \cite{LiangDisplacedNumberSampling}.

The Wigner function is a quasi-probability distribution.  Unlike wave functions it is guaranteed to be real, but unlike classical probabilities, it can be negative. In quantum optics states with negative-valued Wigner functions are defined to be `non-classical.'
It should be noted however that the marginal distributions of momentum and position are always ordinary positive-valued probability distributions (much like the square of the wave function)
\begin{eqnarray}
 \rho_2(P) &=& \int_{-\infty}^{+\infty} dX\, W(X,P)\\
 \rho_1(X)&=&\int_{-\infty}^{+\infty} dP\, W(X,P).
 \label{eq:margqp}
 \end{eqnarray}

 Just as the density matrix contains all the information ever needed to compute the expectation value of any observable via $\langle \hat{\mathcal{O}}\rangle = \mathrm{Tr}\,[\hat{\mathcal{O}}\rho]$, the Wigner function of $\rho$ can be used to obtain the same quantity \shortcite{CahillGlauberPhysRev.177.1882}.

 \begin{exercise}
 From the definition of the Wigner function in App.~\ref{app:Wigner} prove that $$\iint dXdP\, W(X,P)=1.$$
 \end{exercise}

 Since we are dealing with photons in a resonator, it is convenient to replace the position and momentum coordinates of phase space by a single dimensionless complex number $\beta$ that expresses both position and momentum in units of `square root of photons.'
 %
 The displacement operator is then given by
 \begin{equation}
 \mathcal{D}(\beta) = e^{\beta a^\dagger - \beta^* a}.
 \label{eq:displacea}
 \end{equation}
 We can check this expression by noting that
 \begin{equation}
 \mathcal{D}(-\beta) a \mathcal{D}(+\beta) = a + \beta
 \end{equation}
 and that the displacement of the vacuum yields the corresponding coherent state
 \begin{equation}
 \mathcal{D}(\beta)|0\rangle = e^{\beta a^\dagger}e^{-\beta^* a} e^{-\frac{1}{2}[\beta a^\dagger,-\beta^* a]}|0\rangle
 =e^{-\frac{|\beta|^2}{2}}e^{\beta a^\dagger}|0\rangle=|\beta\rangle
 \label{eq:displacedvac}
 \end{equation}
 with mean photon number $\bar n=|\beta|^2$.  In these units, the (now-dimensionless) Wigner function is given by
 \begin{equation}
 W(\beta) = \frac{2}{\pi}\,\mathrm{Tr}\,\left\{\rho {\mathcal D}^\dagger(-\beta) \hat \Pi{\mathcal D}(-\beta)\right\},
 \label{eq:Walpha}
 \end{equation}
 where we have used the fact that the Jacobian for the transformation from $X,P$ to $\beta=\beta_\mathrm{R}+i\beta_\mathrm{I}$ obeys \begin{equation}
 dX\,dP=2\hbar\, d\beta_\mathrm{R}\,d\beta_\mathrm{I}.
 \label{eq:Jacobian}
 \end{equation}

\begin{exercise}
Verify eqn~\ref{eq:displacea}.  Hint: differentiate both sides of the equation with respect to (the magnitude of) $\beta$.
\end{exercise}
\begin{exercise}  a) Verify eqn~\ref{eq:Jacobian}.
b) Verify the normalization  $\iint d\beta_\mathrm{R}\,d\beta_\mathrm{I}\, W(\beta)=1$ directly from eqn~\ref{eq:Walpha}.
\end{exercise}
\begin{exercise}
Using eqn~\ref{eq:displacea}, show that the Wigner function of the coherent state $|\alpha\rangle$ is a Gaussian centered at the point $\alpha$ and given by:
$W(\beta)=  \frac{2}{\pi}e^{-2|\alpha-\beta|^2}.$
\end{exercise}

As described in App.~\ref{app:Wigner}, the experimental procedure to measure the Wigner function is very simple.  One applies a microwave tone resonant with the cavity and having the appropriate phase, amplitude and duration to displace the cavity in phase space by the desired amount.  One then uses the standard procedure described above to determine whether the parity is $+1$ or $-1$.  This procedure is repeated and the results averaged to obtain the mean value of the parity operator.   This `continuous variable' quantum tomography procedure is vastly simpler and more accurate than having to measure different combinations of multi-qubit correlators as required to do tomography in the discrete variable case.   Furthermore the microwave cavity is simple (literally an `empty box'), and the tomography can be performed with only a single ancilla transmon and cavity.   This hardware efficiency is extremely powerful.

\section{Creating Cats}
The Schr\"odinger Cat paradox has a long and storied history in quantum mechanics.  In circuit QED the cat being dead or alive is represented by a qubit being in the ground state $|g\rangle$ or the excited state $|e\rangle$.  The role of the poison molecules in the air is played by a coherent state $|\alpha\rangle$ of photons in the cavity.  The quantum state we want to create is
\begin{equation}
|\Psi\rangle = \frac{1}{\sqrt{2}}[|e\rangle|0\rangle\pm |g\rangle|\alpha\rangle.
\label{eq:SchCat}
\end{equation}
This is a coherent superposition of `cat alive, no poison in the air' and `cat dead, poison in the air.'  Notice that this is \emph{not} a superpostion of `cat dead' and `cat alive.' (At this point the choice $\pm1$ of the phase of superposition is not particularly important, but we retain it for later use.)  It is an \emph{entangled} state between the cat and the poison.  For large $\alpha$ the two states of the cavity are macroscopically distinct and orthogonal.  Note that it is important that the cavity be long-lived because the (initial) rate of photons leaking out from the coherent state is $\kappa  \bar n=\kappa |\alpha|^2$, where $\kappa$ is the cavity damping rate.  If a photon ever leaks out of the cavity into the environment, the environment immediately collapses the state to a product state in which the cat is dead. On the other hand, if no photons are detected after a long time, we grow more and more certain that they were never there in the first place and the second component of the state gradually damps out collapsing the system to $|e\rangle|0\rangle$. This quantum back action from \emph{not} observing a photon entering the environment is quite a novel and subtle effect.  See \shortcite{HarocheRaimondcQEDBook,KittenCodes} for further discussion.\footnote{ See also the short story `Silver Blaze,' by Sir Arthur Conan Doyle in which Sherlock Holmes solves a crime by noting the curious fact that a dog did not bark in the night.}

How do we create this Schr\"odinger cat state?  It is surprisingly easy in the strong-dispersive coupling regime.  The recipe is as follows.  First apply a $\pi/2$ pulse to the qubit to put it in the state $|+x\rangle=\frac{1}{\sqrt{2}}[|g\rangle+|e\rangle]$.  Then apply a drive tone to the cavity at frequency $\omega_\mathrm{c}$.  As we can see from Fig.~\ref{fig:StrongDispersiveCavityShift}, this drive will be on resonance with the cavity (and hence able to displace the cavity state from $|0\rangle$ to $|\alpha\rangle$) if and only if the qubit is the ground state. If the qubit is in the excited state, the cavity remains in the vacuum state (to a very good approximation since the drive is thousands of linewidths off resonance in this case).  Hence we immediately obtain the cat of Eqn~\ref{eq:SchCat}.  It is straightforward to produce very `large' cats with $\bar n=|\alpha|^2$ corresponding to hundreds of photons but our ability to do high-fidelity state tomography begins to degrade above about 100 photons \shortcite{100photonCat}.

\begin{figure}[htb]        
\begin{center}              
 \includegraphics[width=4.0in]{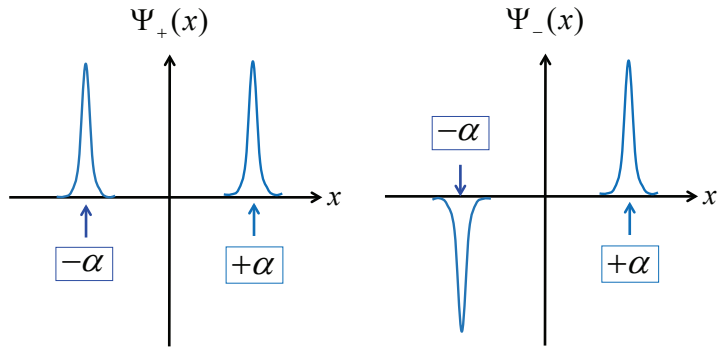}
\end{center}
\caption{First quantization wave functions for even and odd parity cat states.
}
\label{fig:catstategauss}    
\end{figure}

Another interesting state, confusingly referred to in the literature not as a `Schr\"odinger cat' but as a `cat state' of photons is given by
\begin{equation}
|\Psi_\pm\rangle = |g\rangle\frac{1}{\sqrt{2}}[|+\alpha\rangle \pm |-\alpha\rangle].
\label{eq:catstatephotons}
\end{equation}
This is a product state in which the qubit is not entangled with the cavity but the cavity is in a quantum superposition of two different (and for large $\alpha$, macroscopically distinct) coherent states. (The normalization is approximate and only becomes exact in the asymptotic limit of large $|\alpha|$.  We will ignore this detail throughout our discussion.) Because the qubit state factors out we will drop it from further discussion. The $\pm$ sign in the superposition labels the photon number parity of the state.  That is, these are eigenstates of the parity operator:
\begin{equation}
\hat\Pi |\Psi_\pm\rangle =\pm |\Psi_\pm\rangle.
\end{equation}
The parity of the state can be directly verified using the defining property of coherent states in eqn~\ref{eq:displacedvac} but is perhaps best visualzed in terms of the first-quantization wave function for the cat state.  Since the ground state wave function is a Gaussian, the cat state consists of a sum or difference of two displaced Gaussians as illustrated in Fig.~\ref{fig:catstategauss}.  We see that the spatial parity symmetry under coordinate reflection and the photon number parity are entirely equivalent.  Mathematically this is because photon number parity reverses the position and momentum of the oscillator (i.e.\ inverts phase space), a fact which is readily verified using, e.g. $\hat\Pi \hat x\hat\Pi\sim\hat\Pi (a+a^\dagger)\hat\Pi=-(a+a^\dagger)$.

\begin{figure}[htb]        
\begin{center}              
\includegraphics[width=2.0in]{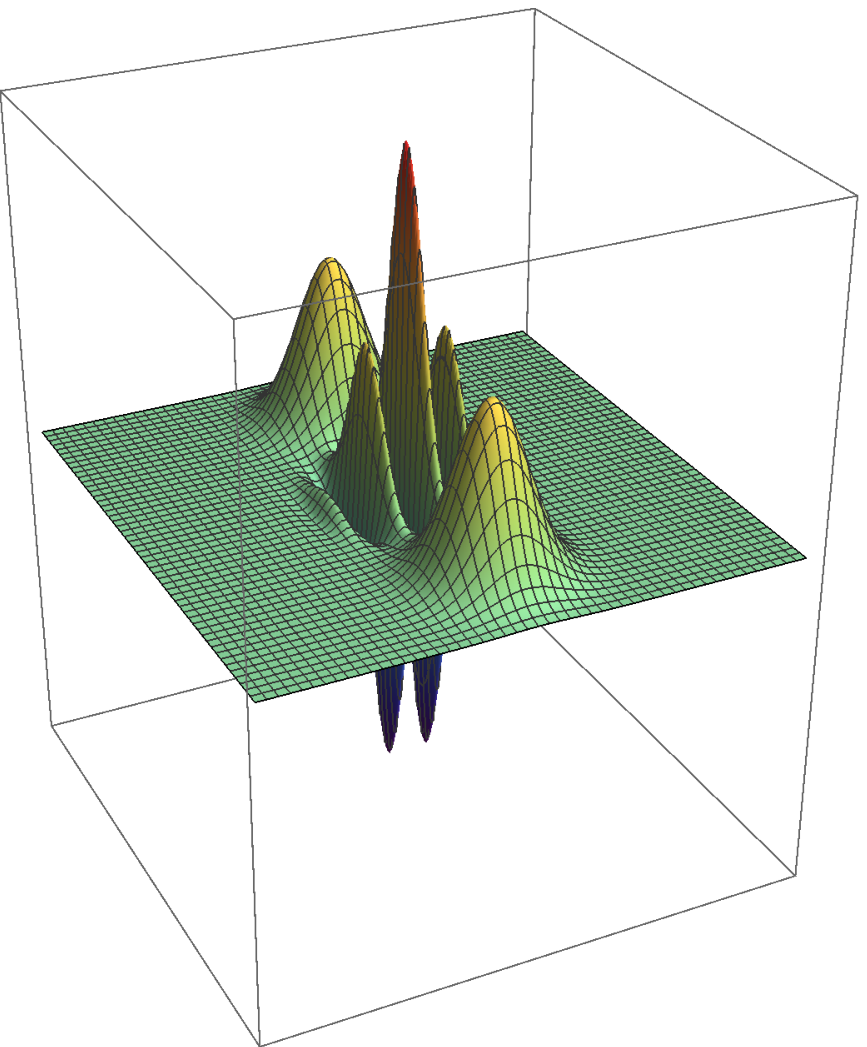}
\includegraphics[width=2.0in]{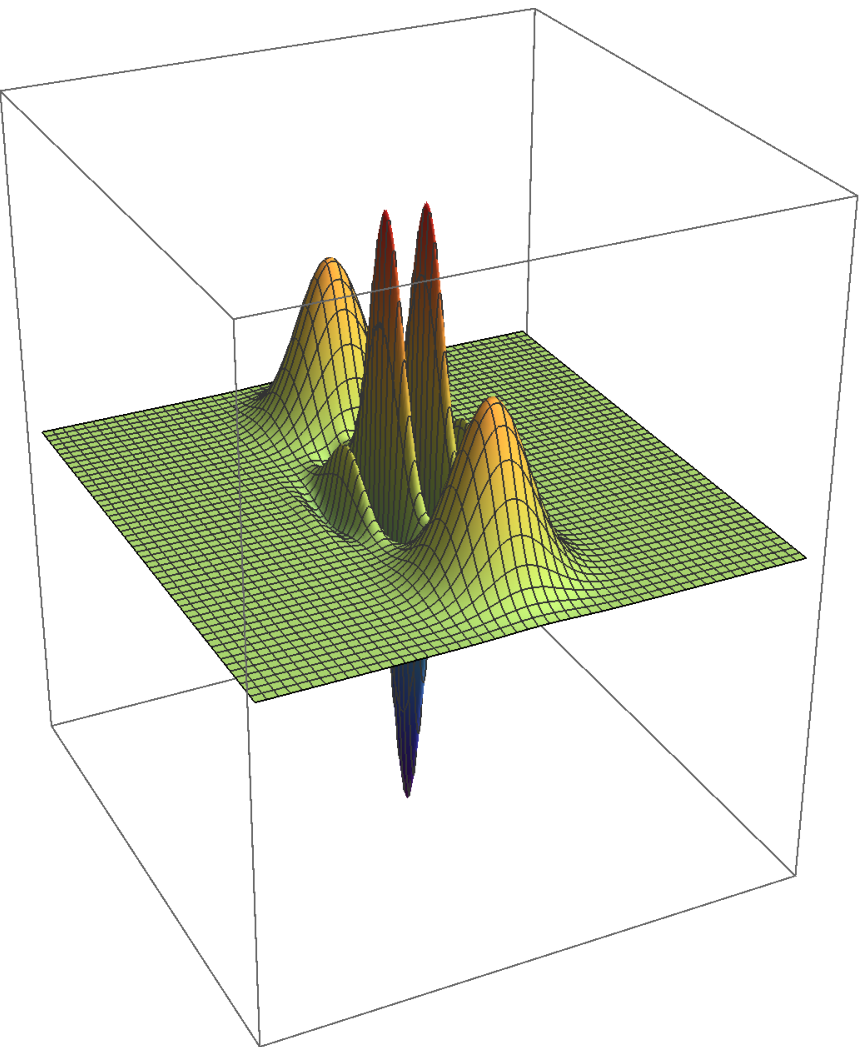}
\end{center}
\caption{Wigner functions for even- (left panel) and odd-parity (right panel) cat states of size $\alpha=2.5$. The foreground and background peaks are associated with the coherent states $|\pm\alpha\rangle$. Note that the parity oscillations near the origin have opposite sign and represent interference fringes associated with the coherence of the superposition of the two distinct states.
}
\label{fig:catstateWigner}    
\end{figure}

The Wigner function of cat states of photons is very interesting.  As one can see in fig.~\ref{fig:catstateWigner}, there are peaks of quasi-probability in phase space at $\pm \alpha$ as expected, but in addition the cat has `whiskers.'  These periodic oscillations are a kind of interference pattern (much like a two-slit interference pattern between the two lobes) that are present for cat states but not for incoherence mixtures of $|+\alpha\rangle$ and $|-\alpha\rangle$.

\begin{exercise}
Using eqn~\ref{eq:Walpha} to write down an (approximate) analytic expression for the Wigner function $W(\beta)$ of a cat state.  Assume an even (odd) cat of the form in eqn~\ref{eq:catstatephotons} with $\alpha$ real and positive, and sufficiently large that the normalization of the states is well-approximated by the $\sqrt{2}$ factor in eqn~\ref{eq:catstatephotons}.  Show that in this limit
\begin{equation}
W(\beta)=\frac{2}{\pi} e^{-2|\beta|^2}\left\{\pm\cos[4\alpha (\mathrm{Im}\beta)]+\cosh[4\alpha(\mathrm{Re}\beta)]e^{-2\alpha^2}\right\}.
\end{equation}
Using this result show that there are no interference fringes in the Wigner function of an incoherent mixture of an even and an odd cat.
\end{exercise}

There is a simple recipe for deterministically creating cat states of photons begins with the Schr\"odinger cat of eqn~\ref{eq:SchCat} except that the coherent state amplitude should be $2\alpha$ instead of $\alpha$.  The trick is to figure out how to disentangle the qubit from the cavity. From eqn~\ref{eq:SchCat} we see that we need to be able to flip the qubit if an only the cavity is in the vacuum state.  When the cavity is in the state $|2\alpha\rangle$ it has (for large $|\alpha|$) negligible amplitude to have zero photons. We can carry out this special disentangling operation again using the strong-dispersive coupling.  We rely on the quantized light shift of the qubit transition frequency by applying a $\pi$ pulse to the qubit at frequency $\omega_\mathrm{q}$ which is the transition frequency of the qubit when there are zero photons in the cavity.  This `number selective $\pi$ pulse yields the product state
\begin{equation}
|\Psi'\rangle = |g\rangle\frac{1}{\sqrt{2}}[|0\rangle \pm |2\alpha\rangle].
\end{equation}
The final step is simply to apply a drive to displace the cavity by a distance $-\alpha$ in phase space to produce the cat state shown in eqn~\ref{eq:catstatephotons}.  The creation of this cat can be verified via quantum state tomography via measuring the Wigner function as described above.  One can also use quantum jump spectroscopy to find the photon number distribution and see that even (odd) cats contain only even (odd) numbers of photons.  For the case of circuit QED, both of these checks have been carried out in Ref.~\shortcite{Vlastakisbigcat} with negative Wigner function fringes measured in states with size up to $d^2=111$ photons, where $d=2\alpha$ is the distance between the two coherent states of the cat.

 Large cat states can be readily produced for microwave photons in circuit QED \shortcite{100photonCat} and cavity QED with Rydberg atoms \shortcite{RevModPhys.85.1083} and produced in the phonon modes of trapped ions using spin-dependent optical forces acting on the ion motional degree of freedom \shortcite{Monroe1996,PhysRevLett.98.063603,PhysRevLett.105.263602,RevModPhys.85.1103,PhysRevLett.116.140402}.

\begin{figure}[htb]        
\begin{center}              
\includegraphics[width=4.0in]{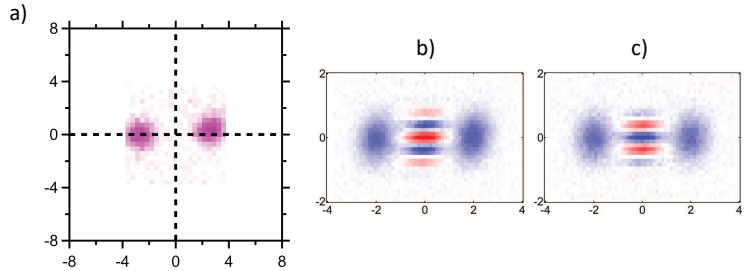}
\end{center}
\caption{Non-deterministic production of a cat state via the back action of a parity measurement of a coherent state with amplitude $\alpha=2$ ($\bar n=4$). a) Wigner function of the incoherent mixture state resulting from tracing over the outcome of the parity measurement; b) Wigner function when the outcome of the parity measurement is odd; c) Wigner function when the outcome of the parity measurement is even.  The phase of the fringe oscillations is opposite in the even and odd cats. Adapted from 
Sun et al.~(2014).
}
\label{fig:nondetermcat}    
\end{figure}

 There is another interesting recipe for producing cat states.  While simple, this recipe produces cats with non-deterministic parity.  Start with a coherent state $|\alpha\rangle$ and write it as a coherent superposition of an even and an odd cat (again assuming $|\alpha|$ is large for simplicity)
 \begin{equation}
 |\alpha\rangle = \frac{1}{\sqrt{2}}[|\Psi_+\rangle + |\Psi_-\rangle],
 \end{equation}
 where $|\Psi_\pm\rangle$ is given by eqn~\ref{eq:catstatephotons} (except we continue to drop the qubit state since it factors out).
Now simply follow the parity measurement protocol described above.  This collapses the state onto definite (but random) parity \shortcite{Brune_PhysRevA.45.5193,Sun14} and hence the measurement back action creates a cat state!  Fig.~\ref{fig:nondetermcat} shows the Wigner function for this process conditioned on the state of the qubit used to perform the parity measurement \shortcite{Sun14}.

 \begin{exercise}
 For large $|\alpha|$ the non-deterministic procedure above produces even and odd cats randomly but with equal probability.  Show that for small $|\alpha|$ there is a bias towards producing even cats more frequently than odd cats.  Compute the probability.  (Hint: in the limit $|\alpha|\rightarrow 0$, we obtain the vacuum state which is definitely even parity.)
 \end{exercise}

 \section{Decoherence of Cat States of Photons}
 For the case of physical objects like trapped ions, the cat state splits the ion wave packet into two positions which can be separated by an amount much larger than their zero-point position uncertainty.   If environmental degrees of freedom are able to gain information about the position of the ions then of course the superposition collapses to a single coherent state.  For example, a stray gas atom might bounce off one of the trapped ions or the electric field from the moving ions may induce damping via energy transfer into nearby metallic structures. For the electromagnetic oscillator the `position' of the `object' that is oscillating is the electric field strength (say) at some selected point inside the cavity. (That point can be selected arbitrarily as long as it is not a nodal point of the mode.)  Because the frequency of the cavity is set by geometry it is very stable and there is very little dephasing of the electromagnetic oscillations.  The primary interaction with the environment is through weak damping of the resonator via the port that brings in the drive tones.  When a photon leaks out of the cavity the parity of the cat must of course change.  Equivalently this follows from the property of coherent states that they are eigenstates of the destruction operator:
\begin{eqnarray}
a|+\alpha\rangle &=&(+\alpha)\,|+\alpha\rangle\\
a|-\alpha\rangle &=&(-\alpha)\,|-\alpha\rangle
\end{eqnarray}
Interestingly this means that photon loss does not dephase a coherent state (indeed it does nothing to a coherent state!) but photon loss is a dephasing error on cat states.    Since the rate of photon loss $\kappa |\alpha|^2$ grows with the `size' of the cat, macroscopic cats dephase quickly and become classical mixtures of two coherent states.

If photon loss does nothing to a coherent state how does its energy decay?  This comes from the back action associated with the intervals in which photons are not observed leaking out that was mentioned above.   This leads to a deterministic decay of the amplitude
\begin{equation}
|\alpha(t)\rangle = |e^{-\frac{\kappa}{2}t}\alpha(0)\rangle.
\end{equation}
One remakrable feature of this is that even the cavity is emitting photons into the environment, it remains in a pure state and does \emph{not} become entangled with the environment. This is a unique feature of coherent states in simple harmonic oscillators \shortcite{HarocheRaimondcQEDBook}.

Let us now move beyond simple coherent states to more general (and possibly mixed) states of the damped oscillator. The time evolution of the density matrix can be solved exactly by considering all the possible quantum trajectories specified by the random instants in time that photons leak out of the cavity \shortcite{HarocheRaimondcQEDBook,KittenCodes}.  The so-called Kraus operators that describe the CPTP (completely positive trace-preserving) map can be organized according to the total number of photons lost in time $t$.  In the absence of any Kerr non-linearities, the actual time that the photons are lost has no effect on the time evolution:
\begin{equation}
\rho(t) = \sum_{\ell=0}^\infty {\hat E}_\ell \rho(0) \hat E^\dagger_\ell,
\label{eq:CPTP}
\end{equation}
where the $\ell$th Kraus operator describing the loss of $\ell$ photons is
\begin{equation}
\hat E_\ell(t) = \sqrt{\frac{(1-e^{-\kappa {t}})^\ell}{\ell !}}e^{-\frac{\kappa {t}}{2}\hat n}  a^\ell.
\label{eq:kraus}
\end{equation}

\begin{exercise}
Prove that the CPTP in eqn~\ref{eq:CPTP} is in fact trace-preserving by showing that $$\sum_{\ell=0}^\infty \hat E^\dagger_\ell \hat E_\ell=\hat I,$$ where $\hat I$ is the identity operator.  Hint:  Prove the identity is true for an arbitrary Fock state $|n\rangle$ and use the fact that the Fock states are complete.
\end{exercise}

\begin{exercise}
Compute the expectation value of the parity over time as a damped oscillator decays starting from: a) an initial even cat state, and b) an initial odd cat state.  The expectation value of the parity will begin at $\pm 1$, then (for large cats) decay close to zero  and then end up at $+1$ because the vacuum is even parity.
\end{exercise}

\begin{exercise}
Compute the Wigner function over time for a damped oscillator decaying from: a) an initial even cat state, and b) an initial odd cat state.
\end{exercise}

\subsection{Quantum Error Correction Using Cat States}
Cat states are well-known to be delicate and their phase coherence is notoriously subject to rapid decay due to photon loss.  It therefore seems like a very bad idea to try to use them to store quantum information.  In fact the opposite is true. Refs.~\shortcite{Leghtas13,Mirrahimi14}   propose and Ref.~\shortcite{Nissim16} demonstrates a clever scheme for encoding quantum information in two (nearly) orthogonal code words consisting of even-parity cat states
\begin{eqnarray}
|W_1\rangle &=& \frac{1}{\sqrt{2}}[|+\alpha\rangle +|-\alpha\rangle]\\
|W_2\rangle &=& \frac{1}{\sqrt{2}}[|+i\alpha\rangle +|-i\alpha\rangle],
\end{eqnarray}
where (without loss of generality) $\alpha$ is real and positive.  (We assume $\alpha$ is large enough that the states are nearly orthonormal.)  These states are indeed highly sensitive to photon loss and quickly dephase.  However as noted above, the source of the incoherence is the loss of parity information when we do not know how many photons have leaked out of the resonator.  Fortunately, this problem can be overcome because, as noted above, we have the abiity to make rapid, high fidelity and highly QND measurements of the photon number parity that can be repeated hundreds of times without back action damage to the state.

Let us now examine in detail how by (semi-) continuously monitoring the parity jumps in the system, we can recover the stored quantum information.  Under photon loss the two code words obey the following relations
\begin{eqnarray}
a|W_1\rangle &\rightarrow & \frac{1}{\sqrt{2}}[|{+\alpha}\rangle -|{-\alpha}\rangle]\\
a^2|W_1\rangle &\rightarrow & +|W_1\rangle\\
a|W_2\rangle &\rightarrow & i\frac{1}{\sqrt{2}}[|{+i\alpha}\rangle -|{-i\alpha}\rangle]\\
a^2|W_2\rangle &\rightarrow & -|W_2\rangle.
\end{eqnarray}
After the loss of two photons the parity has returned to being even but we have a phase flip error.  Thus it takes the loss of four photons for an arbitrary superposition of the two code words to return to the original state.  We do not need to correct immediately for each photon loss.  We need only keep track of the total number lost modulo 4 and then conditioned on that, apply one of 4 unitaries to restore the state.  This is extremely powerful and extremely simple and has allowed circuit QED to be the first technology to reach the break-even point for quantum error correction in which the lifetime of the quantum information exceeds that of the best single component of the system \shortcite{Nissim16}.  Assuming perfect parity tracking (and no dephasing of the cavity) the only source of error from parity jumps would be the possibility that the parity jumps more than once in the short interval $\Delta t$ between measurements.  Two jumps would leave the parity untouched and the error monitoring would miss this fact.  If photon loss occurs at rate $\Gamma\sim \kappa \bar n$, then parity monitoring at intervals $\Delta t$ would reduce the error rate from first order $\Gamma$ to second order $\mathcal{O}(\Gamma^2 \Delta t)$.

As noted above, in addition to the photon number jumps which occur at random times, the amplitude of the coherent state decays at rate $\kappa/2$.  Because this is fully deterministic, the logical qubit decoding circuit can simply take this into account.  The code fails at long times because the amplitude $\alpha e^{-\frac{\kappa}{2}t}$ becomes sufficiently small that the two code words are no longer orthonormal.  Novel ideas for `cat pumping' to keep the coherent states energized indefinitely have already been demonstrated experimentally for two-legged cats \shortcite{LeghtasCatPumping,CoherentControlPumpedCat}.    Bosonic codes for quantum error correction are now an active area of research \shortcite{Gottesman01,Terhal15,Chuang97,KittenCodes,CatCommunicationPhysRevLett.119.030502,CodeComparison2017} and will be discussed in detail at a future Les Houches School.

\section{Conclusions and Outlook}
Circuit QED offers access to strong-coupling cavity QED with artificial atoms based on Josephson junctions coupled via antenna elements to microwave photons.   In the dispersive regime where the atom and cavity are strongly detuned from each other, the dispersive coupling between atom and cavity can be several orders of magnitude larger than dissipation rates.   This `strong-dispersive' regime allows robust universal quantum control of the coupled system and permits creation of novel entangled Schr\"odinger cats as well as cat states of photons.   In this regime it is easy to make high-fidelity and highly QND measurements of photon number parity.  This allows production of cat states by measurement back action on coherent states and also allows repeated measurements of the primary error syndrome for cavity decay.  This in turn permits highly hardware-efficient quantum error correction protocols and very simple quantum state tomography through measurement of the Wigner function of the cavity.

In addition to yielding a wonderful new regime to explore cavity QED, these capabilities will be key to a novel quantum computer architecture in which the logical qubits are stored in microwave resonators using bosonic codes and controlled by Josephson-junction based non-linear elements.  Rapid progress towards this goal is being made as evidenced by recent demonstration of a universal gate set for a logical qubit encoded in a cavity \shortcite{HeeresUniversalGateSet}, entangling photon states \shortcite{Wang2011TwoCavityEntanglement} and cat states \shortcite{ChenWang16} between separate cavities, implementation of a CNOT gate between bosonic codes words stored in separate cavities \shortcite{CNOT_LogicalGate} and development of a `catapult' to launch cat states stored in cavity into flying modes \shortcite{SchrodingerCatapult} for use in error correction in quantum communication \shortcite{CatCommunicationPhysRevLett.119.030502} and remote entanglement. 




\appendix

\chapter{The Wigner Function and Displaced Parity Measurements}
\label{app:Wigner}

The density matrix for a quantum system is defined by
 \begin{equation}
 \rho\equiv \sum_j |\psi_j\rangle p_j\langle \psi_j|
 \end{equation}
where where $p_j$ is the statistical probability that the system is found in state $|\psi_j\rangle$.
As we will see further below, this is a useful quantity because it provides all the information needed to calculate the expectation value of any quantum observable $\mathcal O$.
In thermal equilibrium, $|\psi_j\rangle$ is the $j$th energy eigenstate with eigenvalue $\epsilon_j$ and $p_j=\frac{1}{Z}e^{-\beta \epsilon_j}$ is the corresponding Boltzmann weight.  In this case the states are all naturally orthogonal, $\langle \psi_j|\psi_k\rangle = \delta_{jk}$.  It is important to note however that in general the only constraint on the probabilities is that they are non-negative and sum to unity.  Furthermore there is \emph{no} requirement that the states be orthogonal (or complete), only that they be normalized.

The expectation value of an observable ${\mathcal O}$ is given by
\begin{equation}
\langle\langle{\mathcal O}\rangle\rangle=\mathrm{Tr}\,{\mathcal O}\rho
\end{equation}
where the double brackets indicate both quantum and statistical ensemble averages.  To prove this result, let us evaluate the trace in the complete orthonormal set of eigenstates of ${\mathcal O}$ obeying ${\mathcal O}|m\rangle=O_m|m\rangle$.  In this basis the observable has the representation
\begin{equation}
{\mathcal O}=\sum_m |m\rangle O_m\langle m|
\end{equation}
and thus we can write
\begin{eqnarray}
\mathrm{Tr}\,{\mathcal O}\rho &=& \sum_m\langle m|{\mathcal O}\rho|m\rangle\nonumber\\
&=&\sum_m O_m \sum_j \langle m|\psi_j\rangle p_j\langle \psi_j|m\rangle\nonumber\\
&=&\sum_j p_j \sum_m\langle\psi_j|m\rangle O_m \langle m|\psi_j\rangle \nonumber\\
&=&\sum_j p_j \langle\psi_j|{\mathcal O}|\psi_j\rangle \equiv
\langle\langle {\mathcal O}\rangle\rangle.
\end{eqnarray}

The considerations are quite general.  We now specialize to the case of a single-particle system in one spatial dimension.  A useful example is the harmonic oscillator model which might represent a mechanical oscillator or the electromagnetic oscillations of a particular mode of a microwave or optical cavity.  Our first task is to understand the relationship between the quantum density matrix and the classical phase space distribution.

In classical statistical mechanics we are used to thinking about the probability density $P(x,p)$ of finding a particle at a certain point in phase space.  For example, in thermal equilibrium the phase space distribution is simply
  \begin{equation}
  P(x,p)\frac{dxdp}{2\pi\hbar}=\frac{1}{Z}e^{-\beta H(x,p)}\frac{dxdp}{2\pi\hbar},
  \end{equation}
  where the partition function is given by
  \begin{equation}
 Z=\int\int \frac{dxdp}{2\pi\hbar} e^{-\beta H(x,p)},
  \end{equation}
  and where for convenience (and planning ahead for the quantum case) we have made the phase space measure dimensionless by inserting the factor of Planck's constant.
The marginal distributions for position and momentum are found by
\begin{eqnarray}
\rho_1(x)&=&\frac{1}{2\pi\hbar}\int_{-\infty}^{+\infty}dp\,P(x,p)\label{eq:margx0}\\
\rho_2(p)&=&\frac{1}{2\pi\hbar}\int_{-\infty}^{+\infty}dx\,P(x,p).\label{eq:margp0}
\end{eqnarray}

 Things are more complex in quantum mechanics because the observables $\hat x$ and $\hat p$ do not commute and hence cannot be simultaneously measured because of Heisenberg uncertainty. To try to make contact with the classical phase space distribution it is useful to study the quantum density matrix in the position representation given by
\begin{equation}
\rho(x,x')=\langle x|\rho|x'\rangle = \sum_j p_j \psi_j(x)\psi_j^*(x'),
\end{equation}
where the wave functions are given by $\psi_j(x)=\langle x|\psi_j\rangle$.
 It is clear from the Born rule that the marginal distribution for position can be found from the diagonal element of the density matrix
 \begin{equation}
  \rho_1(x)=\rho(x,x)=\sum_j p_j |\psi_j(x)|^2.
 \label{eq:margx}
  \end{equation}
 Likewise the marginal distribution for momentum is given by the diagonal element of the density matrix in the momentum representation
 \begin{equation}
 \rho_2(p)=\langle p|\rho|p\rangle.
 \end{equation}
 We can relate this to the position representation by inserting resolutions of the identity in terms of complete sets of position eigenstates\footnote{Note that we are using unnormalized momentum eigenstates $\langle x|p\rangle = e^{ipx/\hbar}$.  Correspondingly we are not using a factor of the system size $L$ in the integration measure (`density of states in k space') for momentum.}
 \begin{eqnarray}
 \tilde\rho(p,p') &=& \frac{1}{2\pi\hbar}\int_{-\infty}^{+\infty} dxdx'\, \langle p|x\rangle\langle x|\rho|x'\rangle\langle x'|p'\rangle\nonumber\\
 &=& \frac{1}{2\pi\hbar}\int_{-\infty}^{+\infty} dxdx'\, e^{-ipx/\hbar}\rho(x,x')e^{+ip'x'/\hbar}.
 \end{eqnarray}
 Thus the momentum representation of the density matrix is given by the Fourier transform of the position representation.  We see also
 that the marginal distribution for the momentum involves the off-diagonal elements of the real-space density matrix in an essential way
 \begin{equation}
 \rho_2(p) = \frac{1}{2\pi\hbar}\int_{-\infty}^{+\infty} dxdx' e^{-ip(x-x')/\hbar}\rho(x,x').
 \end{equation}
 For later purposes it will be convenient to define `center of mass' and `relative' coordinates
 $y=\frac{x+x'}{2}$ and $\xi=x-x'$ and reexpress this integral as
  \begin{equation}
 \rho_2(p) =\frac{1}{2\pi\hbar} \int_{-\infty}^{+\infty} dy\int_{-\infty}^{+\infty}d\xi\, e^{-ip\xi/\hbar}\rho(y+\xi/2,y-\xi/2).
 \label{eq:margp}
 \end{equation}

 Very early in the history of quantum mechanics, Wigner noticed from the above expression that one could write down a quantity which is a natural extension of the phase space density.  The so-called Wigner `quasi-probability distribution' is defined by
  \begin{equation}
 W(x,p)\equiv\frac{1}{2\pi\hbar} \int_{-\infty}^{+\infty}d\xi\, e^{-ip\xi/\hbar}\rho(x+\xi/2,x-\xi/2).
 \label{eq:Wignerdef}
 \end{equation}
 Using this, eqn~(\ref{eq:margp}) becomes (changing the dummy variable $y$ back to $x$ for notational clarity)
\begin{equation}
 \rho_2(p) = \int_{-\infty}^{+\infty} dx\, W(x,p).
 \label{eq:margp2}
 \end{equation}
 Similarly, by using eqn~(\ref{eq:Wignerdef}), we can write eqn~(\ref{eq:margx}) as
 \begin{equation}
 \rho_1(x)=\int_{-\infty}^{+\infty} dp\, W(x,p).
 \label{eq:margx2}
 \end{equation}
 These equations are analogous to Eqs.~(\ref{eq:margx0},\ref{eq:margp0}) and show that the Wigner distribution is analogous to the classical phase space density $P(x,p)$.  However the fact that position and momentum do not commute turns out to mean that the Wigner distribution need not be positive.  In fact, in quantum optics one often takes the defining characteristic of non-classical states of light to be that they have Wigner distributions which are negative in some regions of phase space.

 The Wigner function is extremely useful in quantum optics because, like the density matrix, it contains complete information about the quantum state of an electromagnetic oscillator mode, but (at least in circuit QED) is much easier to measure.   Through a remarkable mathematical identity \shortcite{Davidovich1,HarocheRaimondcQEDBook} we can relate the Wigner function to the expectation value of the photon number parity, something that can be measured \shortcite{BertetWignerPhysRevLett.89.200402} and is especially easy to measure \shortcite{Vlastakisbigcat} in the strong-coupling regime of circuit QED (a regime not easy to reach in ordinary quantum optics).

 We are used to thinking of the photon number parity operator in its second quantized form
 \begin{equation}
 \hat\Pi=e^{i\pi a^\dagger a}
 \end{equation}
in which its effect on photon Fock states is clear
\begin{equation}
\hat\Pi|n\rangle = (-1)^n|n\rangle,
\end{equation}
and indeed it is in this form that it is easiest to understand how to realize the operation experimentally using time evolution under the cQED qubit-cavity coupling  $\chi \sigma^z a^\dagger a$ in the strong-dispersive limit of large $\chi$ relative to dissipation.  However because the Wigner function has been defined in a first quantization representation in terms of wave functions, it is better here to think about the parity operator in its first-quantized form.  Recalling that the wave functions of the simple harmonic oscillator energy eigenstates alternate in spatial reflection parity as one moves up the ladder, it is clear that photon number parity and spatial reflection parity are one and the same.  That is, if $|x\rangle$ is a position eigensate
\begin{equation}
\hat\Pi|x\rangle =|-x\rangle,
\end{equation}
or equivalently in terms of the wave function
\begin{equation}
\hat\Pi \psi(x) = \langle x|\hat\Pi|\psi\rangle = \psi (-x).
\end{equation}
To further cement the connection, we note that since the position operator is linear in the ladder operators, it is straightforward to verify from the second-quantized representations that
\begin{equation}
\hat\Pi\hat x= - \hat x \hat \Pi.
\end{equation}

We now want to show that we can measure the Wigner function $W(X,P)$ by the following simple and direct recipe \shortcite{Davidovich1}:  (1) displace the oscillator in phase space so that the point $(X,P)$ moves to the origin; (2)  then measure the expectation value of the photon number parity
\begin{equation}
W(X,P)=\frac{1}{\pi\hbar}\mathrm{Tr}\,\left\{{\mathcal D}(-X,-P)\rho {\mathcal D}^\dagger(-X,-P) \hat \Pi\right\},
\label{eq:dispP}
\end{equation}
where $\mathcal D$ is the displacement operator.
Related methods in which one measures not the parity but the full photon number distribution of the displaced state can in principle yield even more robust results in the presence of measurement noise \shortcite{LiangDisplacedNumberSampling}.


Typically in experiment one would make a single `straight-line' displacement. Taking advantage of the fact that $\hat p$ is the generator of displacements in position and $\hat x$ is the generator of displacements in momentum, the `straight-line' displacement operator is given by
\begin{equation}
{\mathcal D}(-X,-P) = e^{-\frac{i}{\hbar}(P\hat x-X\hat p)}.
\label{eq:straightlinedisp}
\end{equation}
In experiment, this displacement operation is readily carried out by simply applying a pulse at the cavity resonance frequency with appropriately chosen amplitude, duration and phase.  In the frame rotating at the cavity frequency the drive corresponds to the following term in the Hamiltonian
\begin{equation}
V(t) = i[\epsilon(t)a^\dagger - \epsilon^*(t) a]
\label{eq:cavitydisplacement}
\end{equation}
where $\epsilon(t)$ is a complex function of time describing the two quadratures of the drive pulse.  The Heisenberg equation of motion
\begin{equation}
\frac{d}{dt}a=i[V(t),a]=\epsilon(t),
\end{equation}
has solution
\begin{equation}
a(t) = a(0) + \int_{-\infty}^t d\tau\, \epsilon(\tau),
\end{equation}
showing that the cavity is simply displaced in phase space by the drive.
  For the `straight-line' displacement discussed above, $\epsilon(t)$ has fixed phase and only the magnitude varies with time.

\begin{exercise}
Find an expression for $\epsilon(t)$ such that time evolution under the drive in eqn~(\ref{eq:cavitydisplacement}) will reproduce eqn~(\ref{eq:straightlinedisp}).  Ignore cavity damping (an assumption which is valid if the pulse duration is short enough).
\end{exercise}

For theoretical convenience in the present calculation, we will carry out the displacement in two steps by using the Feynman disentangling theorem
\begin{equation}
e^{\hat A +\hat B}=e^{\hat A}e^{\hat B}e^{\frac{1}{2}[\hat B,\hat A]}
\end{equation}
(which is valid if $[\hat B,\hat A]$ itself commutes with both $\hat A$ and $\hat B$) to write
\begin{equation}
{\mathcal D}(-X,-P)=e^{i\theta}{\mathcal D}(0,-P){\mathcal D}(-X,0)=e^{i\theta}e^{-\frac{i}{\hbar}P\hat x}e^{+\frac{i}{\hbar}X\hat p},
\end{equation}
where $\theta\equiv {\frac{i}{2\hbar}XP}$.
This form of the expression represents a move of the phase space point $(X,P)$ to the origin by first displacing the system in position by $-X$ and then in momentum by $-P$.  This yields the same final state as the straightline displacement except for an overall phase $\theta$ which arises from the fact that displacements in phase space do not commute.  For present purposes this overall phase drops out and we will ignore it henceforth.

Under this pair of transformations the wave function becomes
\begin{equation}
\psi(x)\rightarrow \psi(x+X)\rightarrow e^{-iPx/\hbar}\psi(x+X).
\end{equation}
More formally, we have two results which will be useful in evaluating eqn~(\ref{eq:dispP})
\begin{eqnarray}
\langle \xi|{\mathcal D}(0,-P){\mathcal D}(-X,0)|\psi \rangle &=& e^{-iP\xi/\hbar}\psi(\xi+X)\\
\langle \psi|{\mathcal D}^\dagger(-X,0){\mathcal D}^\dagger(0,-P)|\xi\rangle&=&e^{+iP\xi/\hbar}\psi^*(\xi+X).
\end{eqnarray}
Taking the trace in eqn~(\ref{eq:dispP}) in the position basis yields
\begin{eqnarray}
W(X,P)&=&\frac{1}{\pi\hbar}\sum_j p_j\int_{-\infty}^{+\infty}d\xi\, \langle \xi|{\mathcal D}(-X,-P)|\psi_j\rangle  \langle \psi_j| {\mathcal D}^\dagger(-X,-P) \hat \Pi|\xi\rangle\nonumber\\
&=&\frac{1}{\pi\hbar}\sum_j p_j\int_{-\infty}^{+\infty}d\xi\, \langle \xi|{\mathcal D}(-X,-P)|\psi_j\rangle  \langle \psi_j| {\mathcal D}^\dagger(-X,-P) |-\xi\rangle\nonumber\\
&=&\frac{1}{\pi\hbar}\sum_j p_j\int_{-\infty}^{+\infty}d\xi\, e^{-iP2\xi/\hbar}\psi_j(\xi+X)\psi_j^*(-\xi+X)\nonumber\\
&=&\frac{1}{2\pi\hbar}\sum_j p_j\int_{-\infty}^{+\infty}d\xi\, e^{-iP\xi/\hbar}\psi_j(X+\xi/2)\psi_j^*(X-\xi/2),
\label{eq:dispP2}
\end{eqnarray}
which proves that the displaced parity is indeed precisely the Wigner function.


\bibliographystyle{OUPnamed}
\bibliography{Girvin_LesHouches2017_v5}

%
%


\end{document}